\documentclass[12pt]{article}
\usepackage{graphicx}
\usepackage{hyperref}

\newcommand\pubnumber{SNSN-323-63}
\newcommand\pubdate{\today}

\newcommand{\fig}[3]{
  \begin{figure}[htb]
    \centering
    \includegraphics[height=1.5in]{figures/#1}
  \caption{#2} \label{fig:#3}
  \end{figure}
}

\newcommand{\figii}[4]{
  \begin{figure}[htb]
    \centering
    \includegraphics[height=1.5in]{figures/#1}  \hspace{1cm}
    \includegraphics[height=1.5in]{figures/#2} 
  \caption{#3} \label{fig:#4} 
  \end{figure}
}

\def\institute{Universidad de Oviedo, SPAIN}

\def\Title#1{\begin{center} {\Large #1 } \end{center}}
\def\Author#1{\begin{center}{ \sc #1} \end{center}}
\def\Address#1{\begin{center}{ \it #1} \end{center}}

\newcommand\pubblock{\rightline{\begin{tabular}{l} \pubnumber\\
         \pubdate  \end{tabular}}}
\newenvironment{Abstract}{\begin{quotation}  }{\end{quotation}}
\newenvironment{Presented}{\begin{quotation} \begin{center} 
             PRESENTED AT\end{center}\bigskip 
      \begin{center}\begin{large}}{\end{large}\end{center} \end{quotation}}

\newcommand{\cds}[1]{\href{https://cds.cern.ch/record/#1/}{cds.cern.ch/record/#1/}}

\newcommand{\ljets}{$\ell$+jets }

\newcommand{\stt}{$\sigma_{t\bar{t}}~$}

\newcommand{\invfb}{$fb^{-1}~$}
\newcommand{\invpb}{$pb^{-1}~$}
\newcommand{\ttbar}{$\textrm{t}\bar{\textrm{t}}$~}

\newcommand{\sqrts}{$\sqrt{s}$ = }
\newcommand{\arxiv}[1]{\href{https://arxiv.org/abs/#1}{arXiv:#1}}

\newcommand{\xsecfive}{  \sigma_{t\bar{t}}^{pp}(5.02~TeV) = 69.5 \pm 6.1 (stat) \pm 5.6 (syst) \pm 1.6 (lumi)~pb = 69.5 \pm 8.4 (total)~pb }

\newcommand{\xsecdilep}{  \sigma_{t\bar{t}}^{pp}(13~TeV) = 815 \pm 9 (stat) \pm 38 (syst) \pm 19 (lumi)~pb }
\newcommand{\xsecljets}{  \sigma_{t\bar{t}}^{pp}(13~TeV) = 888 \pm 2 (stat) ^{+26}_{-28} (syst) \pm 20 (lumi)~pb}

\newcommand{\xsecseven}{  \sigma_{t\bar{t}}^{pp}(7~TeV) = 173.6 \pm 2.1 (stat) ^{+4.5}_{-4.0} (syst) \pm 3.8 (lumi)~pb}
\newcommand{\xseceight}{  \sigma_{t\bar{t}}^{pp}(8~TeV) = 244.9 \pm 1.4 (stat) ^{+6.3}_{-5.5} (syst) \pm 6.4 (lumi)~pb}

\def\beq{\begin{equation}}
\def\eeq#1{\label{#1}\end{equation}}
\def\eeqn{\end{equation}}

\def\beqa{\begin{eqnarray}}
\def\eeqa#1{\label{#1}\end{eqnarray}}
\def\eeqan{\end{eqnarray}}

\let\bar=\overbar

\def\Dslash{\not{\hbox{\kern-4pt $D$}}}
\def\dslash{\not{\hbox{\kern-2pt $\del$}}}

\def\msb{{\bar{\ssstyle M \kern -1pt S}}}

\begin{document}
\begin{titlepage}
\pubblock

\vfill
\Title{Review of the latest measurements of the inclusive \ttbar production cross section in CMS}
\vfill
\Author{Juan R. Gonz{\'a}lez Fern{\'a}ndez on behalf of the CMS Collaboration}
\Address{\institute}
\vfill
\begin{Abstract}
The measurement of the \ttbar inclusive production cross section (\stt) is crucial to probe QCD predictions, constrain new physic scenarios, proton PDFs, top quark pole mass, $\alpha_{S}$ and many other parameters from the SM. The CMS Collaboration has measured this quantity in different final states and at different centre-of-mass energies, probing QCD calculations and deriving several constraints. In this document a summary of the latest results at each channel and centre-of-mass energy is presented.
\end{Abstract}
\vfill
\begin{Presented}
$10^{th}$ International Workshop on Top Quark Physics\\
Braga, Portugal,  September 17--22, 2017
\end{Presented}
\vfill
\end{titlepage}
\def\thefootnote{\fnsymbol{footnote}}
\setcounter{footnote}{0}

\section{Introduction}
The top quark, the heaviest particle of the SM, is the only quark that decays before hadronizing and its study is decisive to probe QCD predictions. In proton-proton colliders, top quarks are mainly produced in pairs (\ttbar) by gluon-gluon fusion. The measurement of the inclusive top quark pair production cross section can be directly compared with precise QCD predictions at NNLO. This can be used to constrain models beyond the SM and several QCD parameters as proton PDFs, $\alpha_{S}$ or the top quark pole mass. Furthermore, the top quark pair production is a main background in plenty of searches for physics beyond the SM, so the precise measurement of its production cross section is crucial for these searches.

In this document a summary of the latest measurements of the top quark production cross section by the CMS experiment \cite{bib:cms} is presented, at centre-of-mass energies of 7, 8, 13 and 5.02 TeV. 

\section{Legacy measurements at \sqrts 7 and 8 TeV}
Very precise \ttbar measurements have been performed using the full data sets at 7 and 8 TeV recorded by CMS, corresponding to luminosities of 5.0 \invfb and 19.7 \invfb \cite{bib:TOP13004}. Events containing at $e\mu$ pair are selected and classified by categories according to the number of jets and b-tags in the event. The cross section is extracted using a binned likelihood fit to multi-differential final state distributions. The measured cross sections are:

\begin{displaymath}
\xsecseven
\end{displaymath}
\begin{displaymath}
\xseceight
\end{displaymath}

with total uncertainties of 3.6\% and 3.7\% respectively. These precise results have been used to determine the top quark pole mass and to constrain new physics scenarios, in particular the production of the scalar supersymmetric partner of the top quark. These results can be seen in Figure \ref{fig:polemassstop}.

\figii{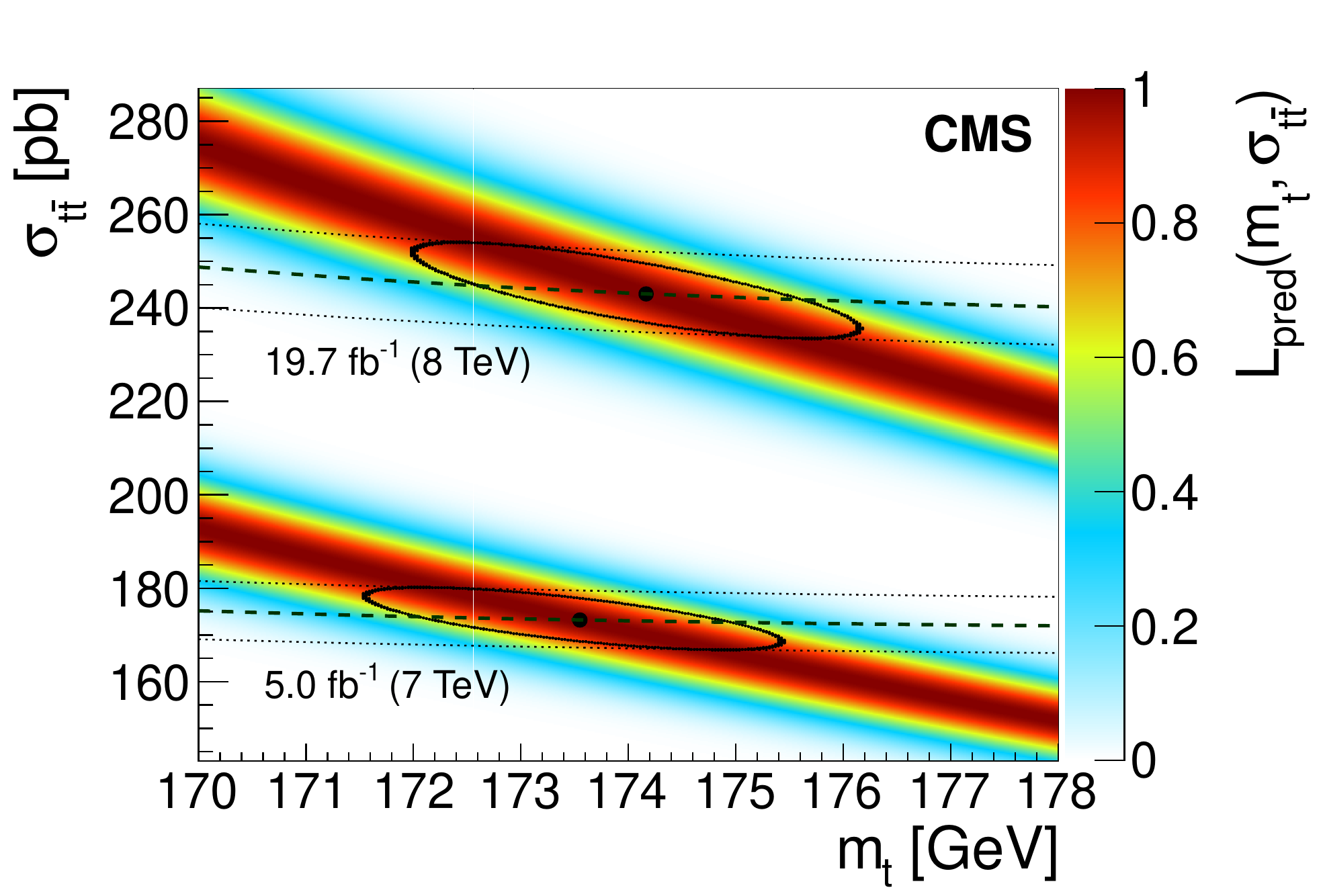}{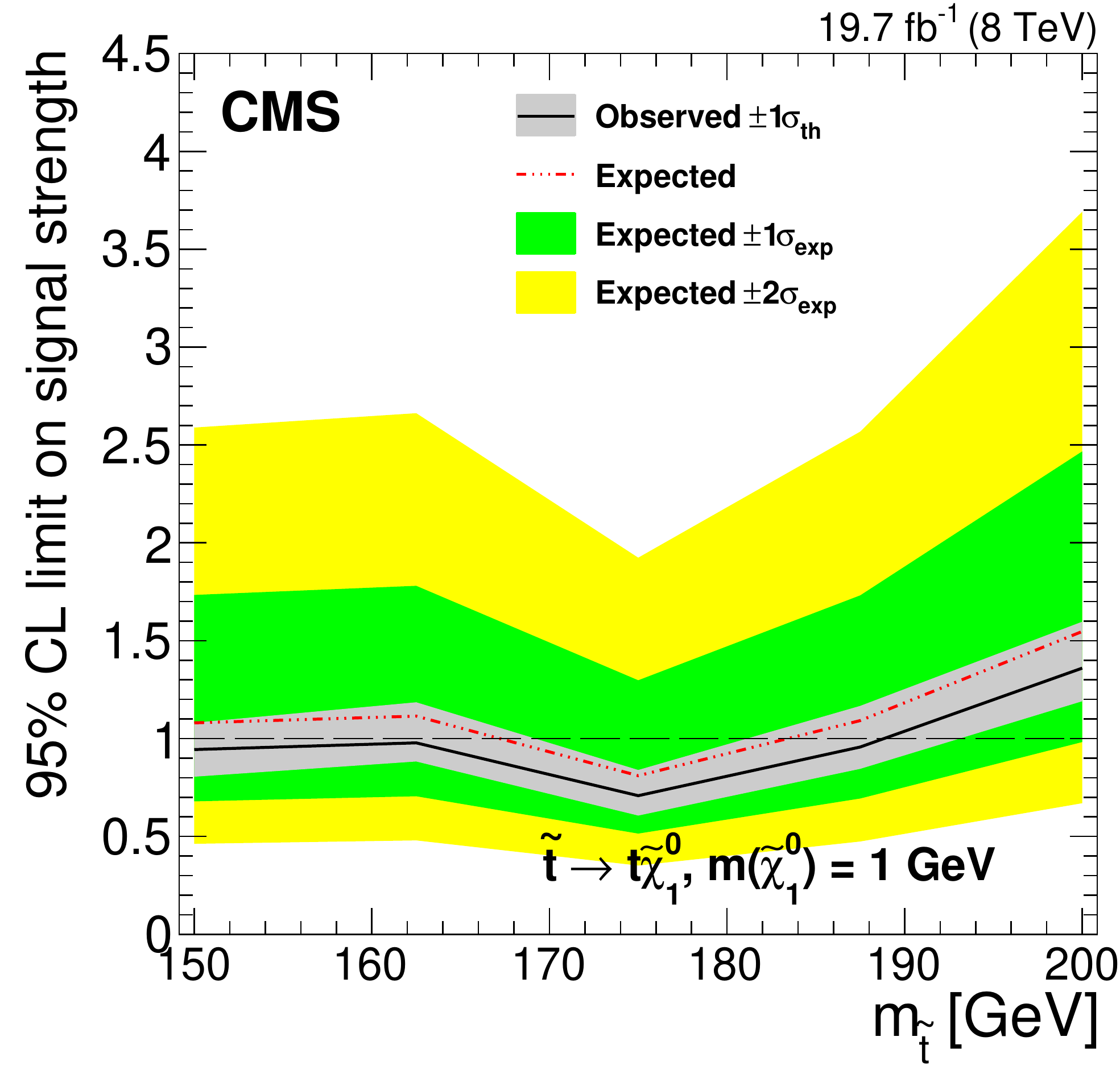}{Likelihood for the predicted dependence of the \ttbar cross section on the top quark pole mass (left) and expected and observed limits at 95\% confidence levels on the signal strength for the stop quark production model with a neutralino mass of 1 GeV, as a function of the stop quark mass \cite{bib:TOP13004}.}{polemassstop}

\section{Latest measurements at \sqrts 13 TeV} 
Since the beginning of the Run 2, several inclusive \ttbar cross section measurements have been performed by CMS. In this section, some highlights of the latest results for each different final state are exposed.

\subsection{Dilepton final state}
A recent measurement using 2.2 \invfb of data \cite{bib:top16005} supersedes the first measurement at this centre-of-mass energy of the \ttbar production cross section by CMS \cite{bib:top15003}. This measurement is performed using events with an electron-muon ($e\mu$) pair with opposite sign The selected $e\mu$ pair mast have an invariant mass of at least 20 GeV. Selected events are required to have at least two reconstructed jets of 30 GeV and $\vert \eta \vert \leq 2.4$ and the presence of at least one b-tagged jet (b-jet).
 
After this selection, the sample is very pure, with background contamination of less than 7\%. This background is mostly estimated from Monte Carlo simulation (MC), with the exception of the so-called NonW/Z leptons background and the Drell-Yan background, which are estimated using data driven techniques, in dedicated control regions. Most of the background processes give very small contributions with the exception of the tW background.

In Figure \ref{fig:dilepxsec} the jet multiplicity is shown after the dilepton selection.

\fig{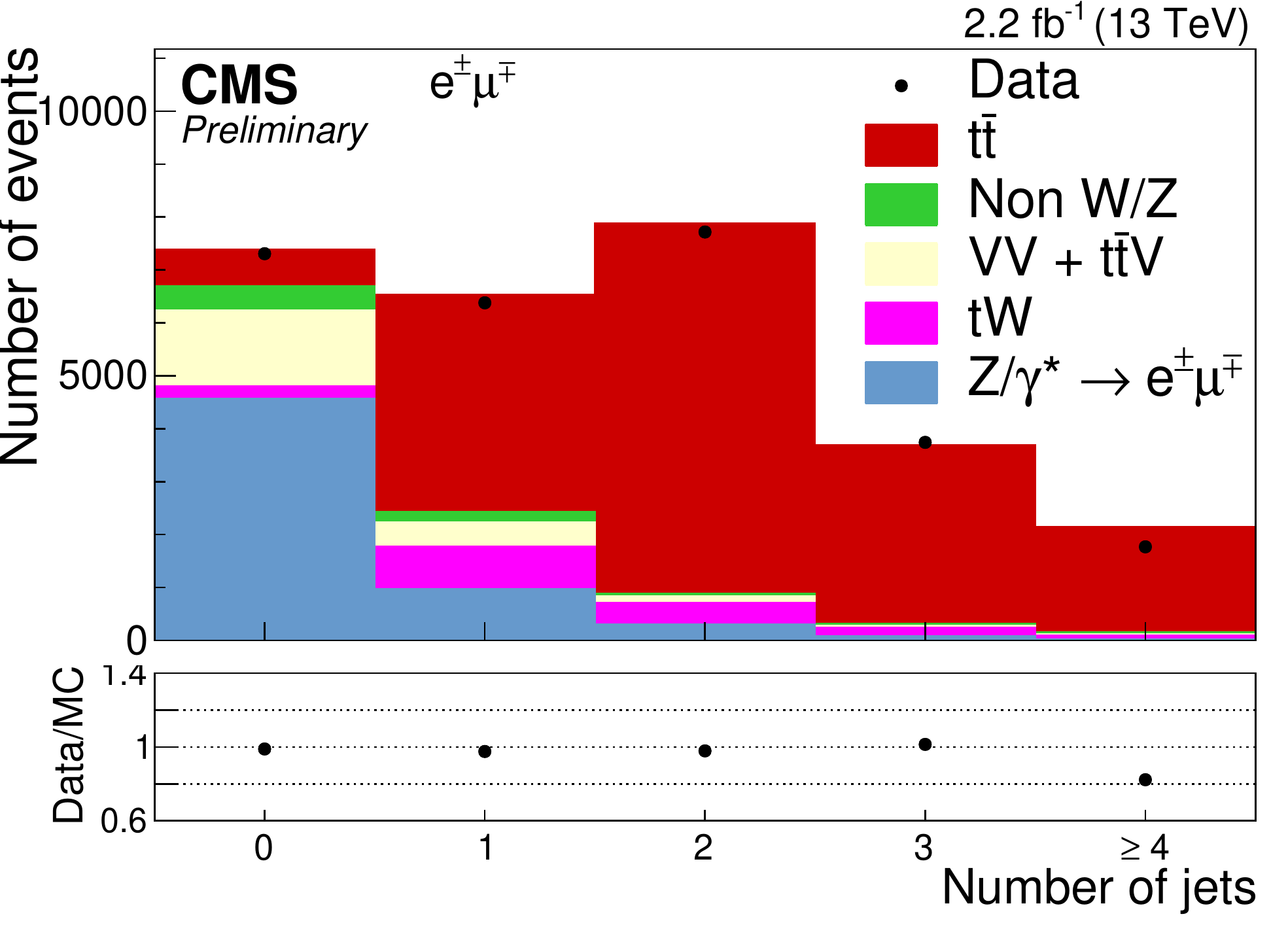}{Distribution of the jet multiplicity for events with one $e^{\pm}\mu^{\mp}$ pair \cite{bib:top16005}.}{dilepxsec}

The measurement is performed using a counting method: background estimation is subtracted from observed data and the result is extrapolated to the full phase space having into account the reconstruction efficiency, the acceptance and the luminosity. Main systematic uncertainties come from \ttbar modeling and jet energy scale. The result is:
\begin{displaymath}
\xsecdilep
\end{displaymath}
with a total relative uncertainty of 5.3\%.

\subsection{\ljets final state}

 A very precise measurement of the \ttbar cross section was recently published by CMS using events with one light charged lepton (e, $\mu$) and jets in the final state \cite{bib:top16006}. Selected events are further categorized by jet and b-jet multiplicities. In this final state, the main backgrounds are W+jets and QCD multijet processes, which are estimated from data in the low-jet and low-b-jet categories. The cross section is extracted by performing a simultaneous binned likelihood fit to all the categories, where systematic uncertainties enter the fit as nuisance parameters and are constrained. 

The largest contributions to the post-fit uncertainty of the measurement come from the uncertainty of the W+jets background estimation, the uncertainty on the luminosity and the \ttbar modeling.

In Figure \ref{fig:lepjetsxsec} the pre-fit numbers of events in each category are shown.

\fig{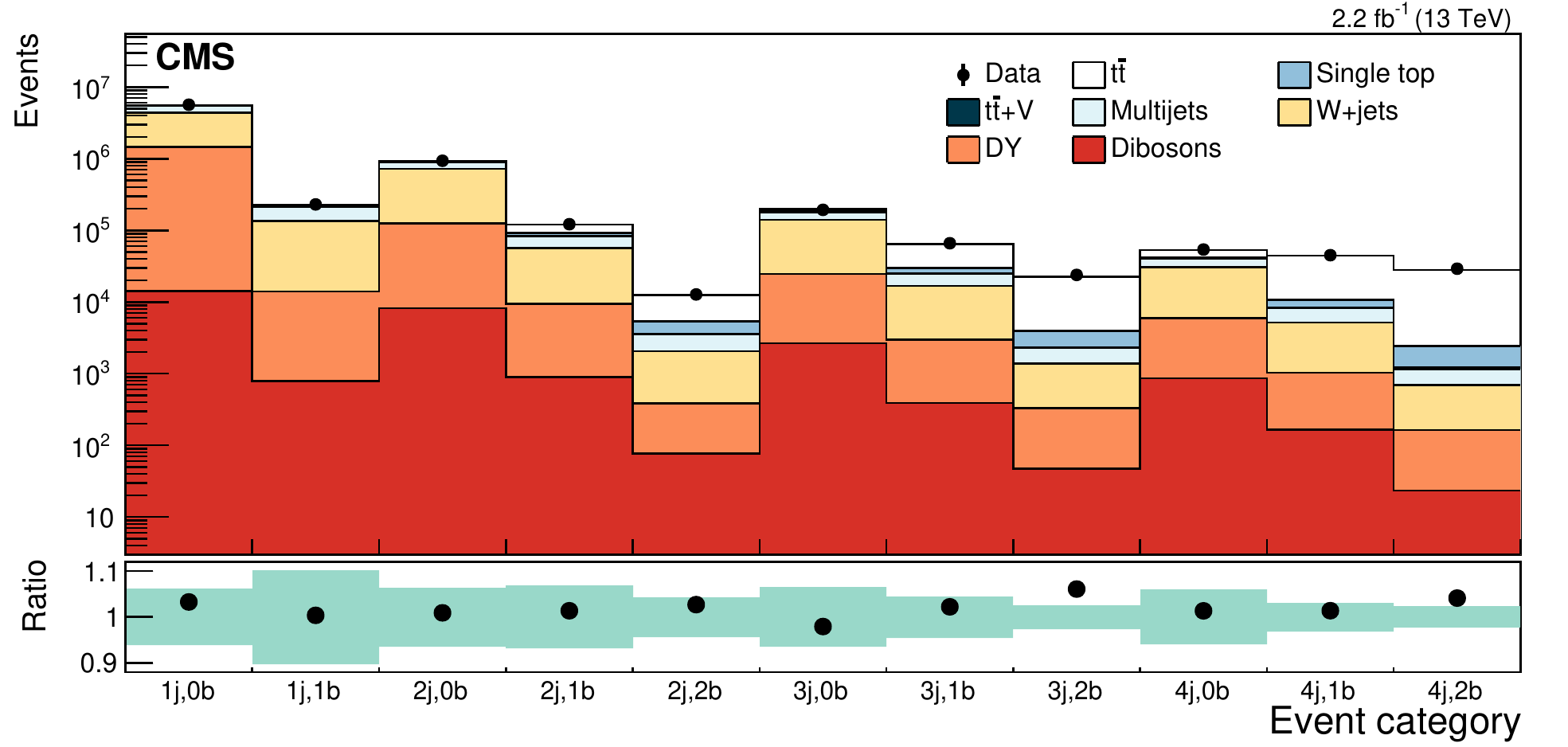}{Yields for data and expected signal and background events for each of the categories of jet (j) and b-tagged jet (b) multiplicity in events with one electron or muon \cite{bib:top16006}.}{lepjetsxsec}

The measured cross section in the \ljets final state is:
\begin{displaymath}
\xsecljets
\end{displaymath}
with a total relative precision of 3.85\%.

\subsection{All hadronic final state}

A recent measurement of the \ttbar cross section was also performed in the full hadronic channel, including a study of the boosted regime \cite{bib:top16013}. Although the large branching fraction in this channel, the hard estimation of the huge QCD background introduces larger uncertainties. The top mass ($m_t$) is reconstructed for the selected events and an unbinned likelihood maximum fit is used to extract the signal strength. 

In Figure \ref{fig:allhad} the $m_t$ distribution is shown.

\fig{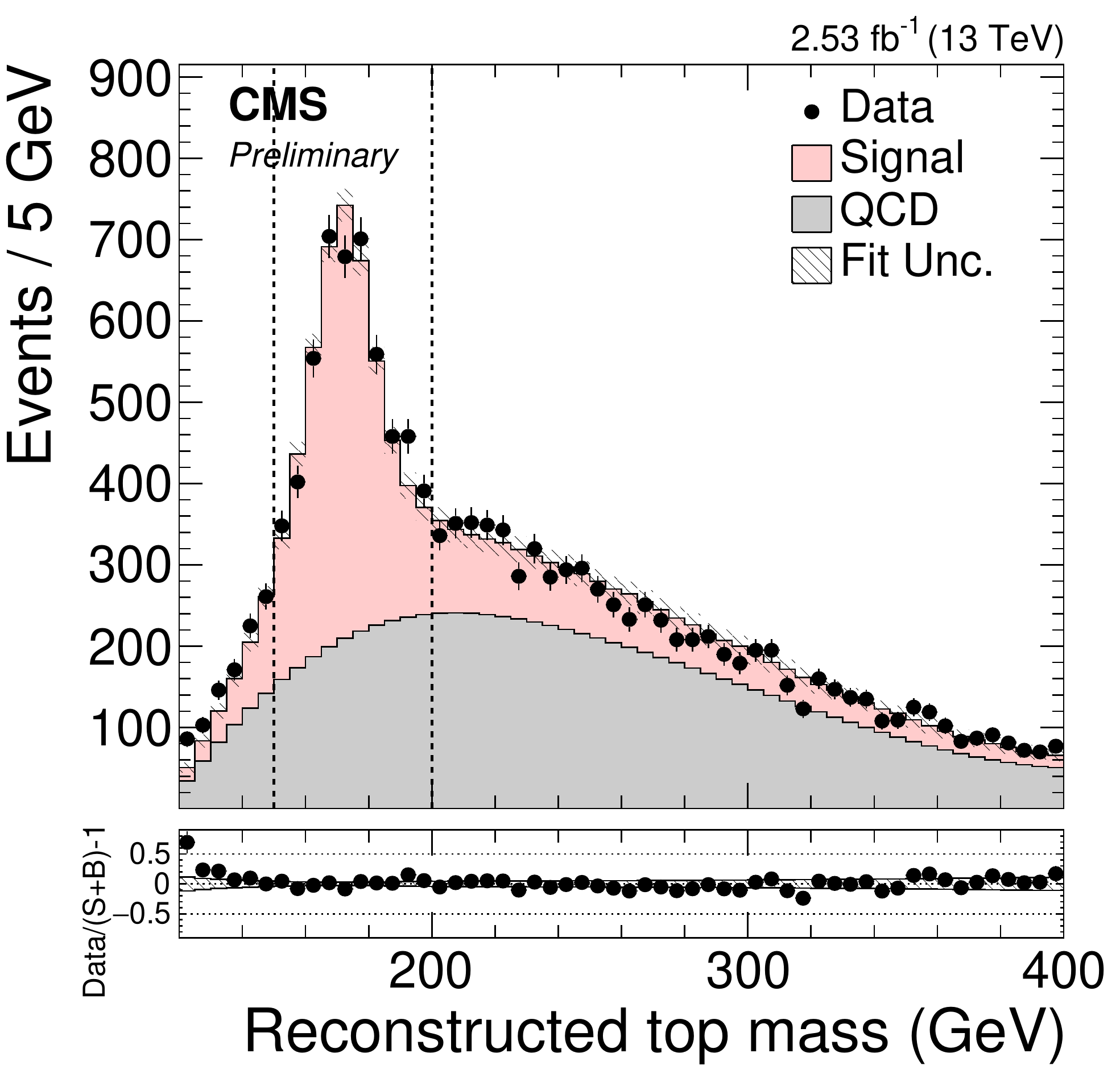}{Reconstructed top mass distribution for selected events in data, expected signal and backgrounds in the all hadronic final state \cite{bib:top16013}.}{allhad}

\section{First measurement at \sqrts 5.02 TeV}

In November 2015, proton-proton collisions were produced by the LHC at a centre-of-mass energy of 5.02 TeV. CMS recorded a total of 27.4 \invpb that were used to measure the \ttbar production cross section at this unexplored energy \cite{bib:top16023}. This measurement was done using events with at least one charged lepton, in several channels. 

A measurement in the \ljets final state was performed using events with one electron or one muon, classified in b-tag multiplicity. The cross section is extracted by a fit to kinematic distributions, in which some background normalization and uncertainties are constrained. Also, measurements in final states with a $\mu\mu$ pair or an $e\mu$ pair were done, using a simple counting method. In all channels, the statistical uncertainty is dominant.

The measurements in the different final states were combined to obtain a more precise result, with a total uncertainty of 12\% and in agreement with SM predictions:

\begin{displaymath}
\xsecfive
\end{displaymath}

This measurement adds a new point in the curve $\sigma_{t\bar{t}}(\sqrt{s})$, which reflects our current knowledge on \ttbar cross sections and the great level of agreement of QCD predictions. Also, at 5 TeV top quark pairs are mostly produced by high-x gluons in proton-proton collisions, so the measurement of \stt is very sensitive to this parameter and can be used to constrain proton PDFs.

\section{Conclusions}

Precision measurements of \ttbar cross sections have been done by CMS in a wide range of centre-of-mass energies, showing a good agreement with NNLO predictions, as shown in Figure \ref{fig:sqrts}.
\fig{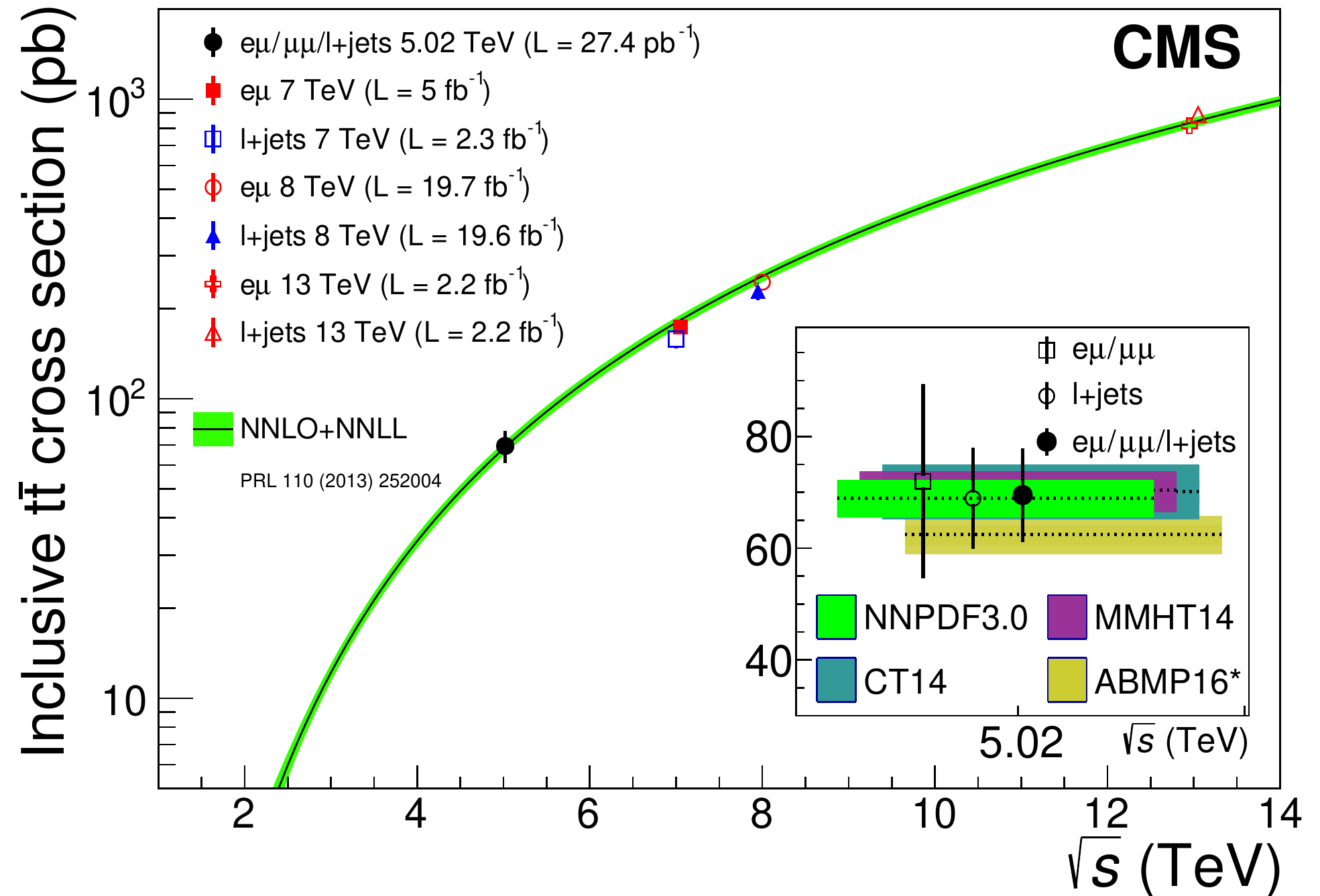}{Summary of \ttbar cross section measurements in CMS for various centre-of-mass energies \cite{bib:top16023}.}{sqrts}


\end{document}